\RequirePackage{fix-cm}
\documentclass[twocolumn]{svjour3}          % twocolumn
\smartqed  % flush right qed marks, e.g. at end of proof
\usepackage{cite}
\usepackage{amsmath,amssymb,amsfonts}
\usepackage{algorithmic}
\usepackage{graphicx}
\usepackage{textcomp}
\usepackage{color}
\usepackage{xcolor}
\usepackage{booktabs}
\usepackage{algorithm} %format of the algorithm 
\usepackage{multirow} %multirow for format of table 
\usepackage{bbm}
\usepackage{stfloats}
\usepackage{mathrsfs}
\usepackage{subfigure}

\begin{document}

\title{Vehicular Fog Computing Enabled  Real-time Collision Warning via Trajectory Calibration 
% \thanks{This work was supported }
\thanks{Corresponding author: Kai Liu}
}


\author{Xincao Xu \and
		Kai Liu \and
		Ke Xiao \and
		Liang Feng \and
		Zhou Wu \and
		Songtao Guo
}

\institute{Xincao Xu \and Kai Liu \and Ke Xiao \and Liang Feng \and Songtao Guo  		\at Key Laboratory of Dependable Service Computing in Cyber Physical Society, Ministry of Education (Chongqing University), and also with the College of Computer Science, Chongqing University, Chongqing, China\\
            \email{near, liukai0807, xiaoke317, liangf, guosongtao@cqu.edu.cn}                   
           \and
           Zhou Wu \at College of Automation, Chongqing University, Chongqing, China\\
           \email{zhouwu@cqu.edu.cn} 
}

\date{Received: date / Accepted: date}

\maketitle

\begin{abstract}
	Vehicular fog computing (VFC) has been envisioned as a promising paradigm for enabling a variety of emerging intelligent transportation systems (ITS). 
	However, due to inevitable as well as non-negligible issues in wireless communication, including transmission latency and packet loss, it is still challenging in implementing safety-critical applications, such as real-time collision warning in vehicular networks.
	In this paper, we present a vehicular fog computing architecture, aiming at supporting effective and real-time collision warning by offloading computation and communication overheads to distributed fog nodes. 
	With the system architecture, we further propose a trajectory calibration based collision warning (TCCW) algorithm along with tailored communication protocols.
	Specifically, an application-layer vehicular-to-infrastructure (V2I) communication delay is fitted by the Stable distribution with real-world field testing data.
	Then, a packet loss detection mechanism is designed.
	Finally, TCCW calibrates real-time vehicle trajectories based on received vehicle status including GPS coordinates, velocity, acceleration, heading direction, as well as the estimation of communication delay and the detection of packet loss.
	For performance evaluation, we build the simulation model and implement conventional solutions including cloud-based warning and fog-based warning without calibration for comparison.
	Real-vehicle trajectories are extracted as the input, and the simulation results demonstrate that the effectiveness of TCCW in terms of the highest precision and recall in a wide range of scenarios.
\end{abstract}

\section{Introduction}
\label{intro}
Vehicular ad hoc networks (VANETs) have attracted great attention to enable a variety of intelligent transportation systems (ITS).
Dedicated short range communication (DSRC) \cite{kenney2011dedicated} is one of the well developed  standards for vehicular communication, where vehicles equipped with on-board units (OBUs) can communicate with both vehicles and roadside units (RSUs) via vehicle-to-vehicle (V2V) and vehicle-to-infrastructure (V2I) communications, respectively.
Meanwhile, another mainstream of vehicular communication standard is based on the development of cellular networks \cite{agiwal2016next}.
The Long-Term Evolution-Vehicle (LTE-V) standard has been developed to enable V2X communication, which is evolving to the 5G based Cellular Vehicle-to-Everything (C-V2X) communication \cite{araniti2013lte}.
A number of studies \cite{ucar2015multihop}\cite{dai2018bandwidth}\cite{ahmed2018cooperative}\cite{liu2019ahierarchical} have investigated on information services in heterogeneous vehicular communication environments. 

In view of the ever-increasing data and computation demands \cite{zhai2018}, the highly dynamic traffic status and network topology \cite{zhai2019}, and the various quality of service (QoS) requirements of different ITS applications \cite{liu2019ahierarchical}, great efforts have been put into the designing of data scheduling algorithms \cite{liu2016cooperative}\cite{dai2016adaptive} \cite{wang2017dynamic}, resource allocation mechanisms \cite{peng2017resource} \cite{ahmed2018secure} and emerging ITS applications \cite{liu2013improving} \cite{dai2016quality}.
In addition, many studies have been focusing on the development of new service paradigms in VANETs based on software defined networking \cite{liu2016cooperative} \cite{wang2019delay} \cite{liu2017coding} and fog computing \cite{liu2019ahierarchical} \cite{hou2016vehicular} \cite{wang2018offloading}, which are envisioned as a promising solutions to enhancing system scalability and flexibility by decoupling control and data planes, as well as enabling low-latency and high-reliability information services by offloading computing, networking, storage, communication, and data resources closer to the end users.
On this basis, different solutions have been proposed for computation offloading \cite{wang2019delay}, distributed service scheduling \cite{sun2018cooperative} and coordinated resource allocation \cite{zhou2019computation} in VANETs.

However, it is still non-trivial to realize safety-critical applications in VANETs due to the stringent real-time requirements on communication and computation.
Besides, intrinsic features such as transmission latency and packet loss in wireless communications are inevitable and also non-negligible for such applications, which make it even challenging for enabling real-time and reliable safety-critical services in VANETs. 

With above motivations, this work presents a vehicular fog computing (VFC) based real-time collision warning architecture in VANETs.
In particular, on the basis of the presented architecture, dedicated solutions are proposed to achieve real-time and accurate collision warning via VFC.
Small-scale system prototype is built in realistic environments to give an insight into the system development. 
Large-scale simulation model is implemented with real vehicle trajectories to evaluate the effectiveness of the proposed architecture and corresponding algorithms.
The main contributions of this work are outlined as follows:
\begin{itemize}
\item We present a VFC based service architecture via DSRC, where RSUs are deployed as fog nodes along the road, which communicate with OBUs mounted on vehicles and process collision warning tasks locally based received vehicle status, including global position system (GPS) coordinates, velocity, acceleration, heading direction, etc. Compared with conventional cloud based system, such a service architecture not only reduces wireless communication delay, but also enhances system scalability and responsiveness by offloading the computation tasks to the distributed fog nodes.
\item We propose a trajectory calibration based collision warning algorithm (TCCW). Specifically, we first derive a packet transmission delay fitting model based on the Stable distribution to estimate the fog based communication latency. The fitting data are obtained by implementing a DSRC based system prototype in realistic vehicular communication environments. Second, we design a packet loss detection mechanism based on the historical information including data transmission frequency and locations of vehicles. Finally, we present detailed procedures of TCCW, which calibrates vehicle trajectories by combining communication latency estimation and  packet loss detection, so as to improves the timeliness and accuracy of the collision warning system.
\item We give a comprehensive performance evaluation. First, we carry out a real-world field testing and obtain the sets of  packet latency in both cloud and fog based systems. Second, we import real-world vehicle trajectories in several selected intersections with different features (e.g., traffic density, vehicle speed, vehicle acceleration) of Cologne, Germany. The simulation results have demonstrated the superiority of TCCW comparing with traditional methods (e.g., cloud-based warning and fog-based warning without trajectory calibration) in terms of both precision and recall on collision warning.
\end{itemize}

The rest of this paper is organized as follows: Section \ref{Related Work} reviews the related work.
Section \ref{System Overview} presents the VFC enabled real-time collision  warning system.
Section \ref{Proposed Algorithm} proposes a trajectory calibration based collision warning algorithm.
Section \ref{Performance Evaluation} builds the simulation model and gives performance evaluation.
Finally, Section \ref{Conclusion} concludes this work.

\section{Related Work}\label{Related Work}

Great efforts have been devoted to improving the performance of VANETs.
DSRC is considered as the de facto standard for vehicular communication.
In America, 75MHz bandwidth of spectrum is allocated by the Federal Communications Commission (FCC) for DSRC.
Based on the allocated spectrum, IEEE published the protocol stack called Wireless Access in Vehicular Environments (WAVE), which consists IEEE 802.11p and IEEE 1609.x.
Although many studies have investigated on vehicular communications via DSRC, most of them only focused on the MAC layer performance.
Yao et al. \cite{yao2013delay} proposed an analytical model to evaluate the broadcast performance under IEEE 802.11p in terms of the mean, deviation and probability distribution of the MAC access delay, and numerical analysis indicated that IEEE 802.11p can provide relatively good performance for higher priority messages.
Zheng et al. \cite{zheng2015performance} analyzed the enhanced distributed channel access mechanism of IEEE 802.11p in terms of the transmission probability, normalized throughput, and average access delay, and simulation results verified the effectiveness of the derived performance models.
Peng et al. \cite{peng2016performance} proposed a general probabilistic performance of multiplatooning communication based on IEEE 802.11p and the simulation results indicated that multiplatooning communications can satisfy the delay requirements of platoon control and on-road safety.

Vehicular fog computing is an emerging paradigm in VANETs to better support low latency, high reliability and large scale ITS applications.
Hou et al. \cite{hou2016vehicular} first presented a viewpoint that vehicles are considered as fog nodes for service provision.
The proposed architecture can better support real-time services and better exploit computation and communication capacities of individual vehicles.
Huang et al. \cite{huang2017vehicular} proposed the vehicular fog computing architecture which comprises three layers, namely, the data generation layer, the fog layer, and the cloud layer.
A fog-assisted traffic control system illustrated the benefits of VFC, and the forensic challenges and potential solutions are discussed.
Wang et al. \cite{wang2018offloading} proposed a fog-enabled real-time traffic management system, aiming to minimize the average response time for events reported by vehicles.
Performance evaluation based on a real-world taxi trajectory dataset demonstrated the effectiveness of the designed method.
Liu et al. \cite{liu2019ahierarchical} proposed a hierarchical system architecture to synthesize the paradigms of software defined networking and fog computing in Internet of Vehicles and comprehensive analysis with respect to the separation of control and data planes, network functions virtualization in heterogeneous resource environments, network slicing for services with different QoS requirements, and the offloading of computation, storage, control, and communication capacities with fog-based services.

Most existing vehicular collision warning systems are based on ranging sensors such as radar or lidar \cite{song2017real} \cite{wu2019series}.
However, they all suffer from non-line-of-sight problems.
With recent advances on computer vision, some studies focused on collision detection based on video camera \cite{wang2016vision} \cite{song2018lane}.
Nevertheless, they may require intensive computation and massive data transmission, which render the system performance on real-time response.  
A few studies considered collision warning via vehicular communications  \cite{hafner2013cooperative} \cite{gelbar2017elastic} \cite{xu2018design}.
Hafner et al. \cite{hafner2013cooperative} leveraged V2V communication technology to implement computationally efficient decentralized algorithms for two-vehicle cooperative collision avoidance at intersections and provided an experimental validation of proposed method.
Gelbar et al. \cite{gelbar2017elastic} proposed a pedestrian collision warning and avoidance system for vehicles based on V2X communication, and demonstrated the effectiveness of the proposed method using hardware-in-the-loop simulations.
Our previous work \cite{xu2018design} designed and implemented a fog computing based collision warning system leveraging V2I communications and demonstrated the superiority of the fog computing based collision warning system in terms of enhancing system responsiveness.

\section{VFC Based Real-time Collision Warning Architecture}\label{System Overview}

\begin{figure}
\centering
  \includegraphics[width=3.0in]{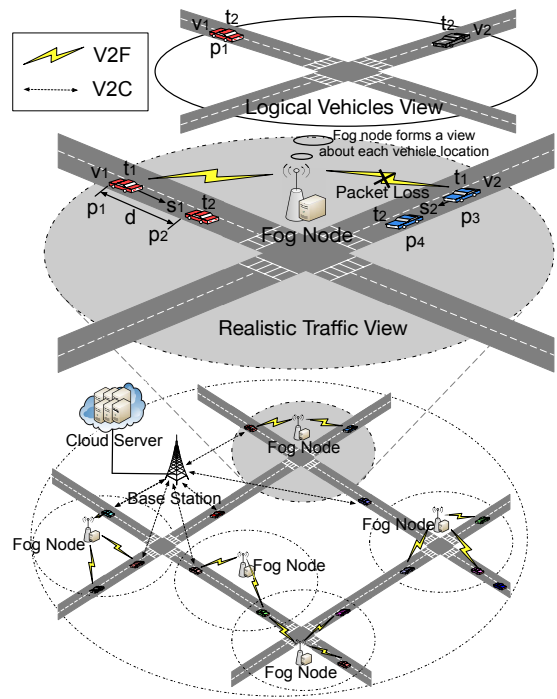}
  \caption{VFC based real-time collision warning architecture}
  \label{fig_1}
\end{figure}

In this section, we propose a VFC based real-time collision warning architecture as shown in Fig. \ref{fig_1}.
In the architecture, communication infrastructures with short radio coverages (e.g., RSUs, 5G small cells) are acted as fog nodes, since they are physically closer to vehicles.
On the other hand, communication infrastructures with wide coverages (e.g., cellular network base stations) are considered as cloud nodes.
Vehicles are able to communicate with fog and cloud nodes via vehicle-to-fog (V2F) and vehicle-to-cloud (V2C) communications, respectively.
Fog nodes installed along the road have certain computation ability.
Cloud nodes are assumed to have unlimited computation capacity, but it may suffer from severe bandwidth competition if all vehicles under its coverage are transmitting data concurrently.  

We take the following scenario as an example to show the procedure of the designed  architecture.
Considering two vehicles (i.e., $v_1$ and $v_2$) are approaching to a road intersection without traffic lights, then they may likely collide, especially if they are blind to each other.
The potential collision can be avoided if the collision warning system, as a typical safety-critical ITS application, is implemented.
Vehicles periodically upload their real-time status including GPS coordinates, velocity, acceleration, heading direction and so on, to the fog node via V2F communication.
However, it is inevitable that the sensed data is inaccurate, such as the inaccuracy of the obtained GPS coordinates due to satellite clock bias, atmospheric delay, and errors in the broadcast ephemeris, etc \cite{liu2013improving}.
Furthermore, the packet loss in wireless communications makes it even challenging for the server to estimate real-time location of moving vehicles.
Therefore, in this presented system, fog nodes will predict vehicle trajectories based on received information and designed algorithms.
Then, a collision detection mechanism is incorporated to determine whether there is potential collisions.  
When a collision is detected based on certain headway threshold, the fog node will send the warning message to corresponding vehicles.

As analyzed above, even though fog node reduces communication delay over traditional centralized cloud computing, it still suffers from inaccurate vehicle information due to inevitable as well as non-negligible issues, including sensor errors, transmission latency and packet loss.
The enlarged part in Fig. \ref{fig_1} shows an example of inaccurate vehicle location in the logical vehicles view of fog node comparing with the realistic traffic view.
Specifically, assume that vehicle $v_1$ is approaching to the intersection with the velocity $s_1$ of $40$ kilometers per hour at time $t_1$, and the position of vehicle $v_1$ is $p_1$.
Meanwhile, vehicle $v_2$ is located at $p_3$ and approaching to the intersection at a speed $s_2$ of $25$ kilometers per hour.
Vehicles send their status to the fog node at time $t_1$ simultaneously, however, the packet, which contains status of vehicle $v_2$, is lost in V2F communication.
Fog node receives vehicle status at time $t_2$ and forms a logical vehicles  view about each vehicle location as shown in the top of Fig. \ref{fig_1}.
Suppose the data size is $500$ kB, which is sufficient for typical ITS applications \cite{liu2013improving}.
DSRC as a typical VANETs communication technology supports data rate from $3$ to $27$ Mb/s \cite{kenney2011dedicated}, where $3$ Mb/s is recommended for transmitting safety-critical messages.
Hence, the time taken to upload vehicle status is around 500 * 8 kb / 3 $\approx$ 1.3 s, namely, ${t_2} - {t_1} \approx 1.3$ $s$.
Vehicle $v_1$ is located at $p_1$ at time $t_2$, and $v_2$ dose not exist in the logical vehicles view of the fog node.
Yet as shown in the realistic traffic view, the realistic positions of vehicles $v_1$ and $v_2$ are $p_2$ and $p_4$, respectively.
The distance error of vehicle $v_1$'s location between the logical vehicles view of fog node and the realistic traffic view is around 40 * 1000 / 3600 m/s * 1.3 $\approx$ 14 m, in other words, $d$ $\approx$ 14 m.
With above example, obviously, it is imperative to design an efficient mechanism based on the presented VFC architecture to enhance system robustness.

\section{Trajectory Calibration Based Collision Warning Algorithm}\label{Proposed Algorithm}

In this section, we propose an algorithm named TCCW, which calibrates vehicle trajectory with both transmission latency estimation and packet loss detection.
First, application-layer V2I communication delay is fitted into the Stable distribution model with real-world field testing data.
Then, we propose a packet loss detection mechanism based on the historical information including data transmission frequency and locations of vehicles.
Finally, we give detailed procedures of TCCW.
Primary notations are listed in Table. \ref{table_notations}.

\begin{table}[ht]\small
\centering
\caption{{\color{blue}Summary of notations}}
\begin{tabular}[t]{ll}
\hline
\hline
Notation&Description\\
\hline
$S$&The Stable distribution\\
$\alpha$&The characteristic exponent of the $S$\\
$\beta$&The skewness parameter of $S$\\
$\mu$&The location parameter of $S$\\
$\sigma$&The scale parameter of $S$\\
$E$&The time slot set\\
${e_{k}}$&One time in set $E$\\
$V$&The vehicle set\\
${v_{i}}$&The $i$th vehicle\\
{\color{blue}
${{l}_{v_{i}}^{e_{r}}}$}&{\color{blue}The location of vehicle ${v_{i}}$ at time $e_r$}\\
{\color{blue}
${{s}_{v_{i}}^{e_{r}}}$}&{\color{blue}The speed of vehicle ${v_{i}}$ at time $e_r$}\\
{\color{blue}
${{a}_{v_{i}}^{e_{r}}}$}&{\color{blue}The acceleration of vehicle ${v_{i}}$ at time $e_r$}\\
{\color{blue}
${{d}_{v_{i}}^{e_{r}}}$}&{\color{blue}The driving direction of vehicle ${v_{i}}$ at time $e_r$}\\
$l_{f}$&The location of the fog node\\
$dis({l_{v_{i}}^{e_{r}}}, {l_{f}})$&The distance between vehicle $v_{i}$ and fog node\\
&at time ${e_r}$\\
$R$&The communication range of DSRC\\
${{M}_{e_k}}$&The received packets at time $e_k$\\
$H_{e_{k}}$&The packets records in fog node at time ${e_{k}}$\\
${{ID}_{e_{k}}}$&The ID set of vehicles at time ${e_{k}}$\\ 
${{ID}_{M_{e_{k}}}}$&The ID set of $M_{e_{b}}$ at time ${e_{k}}$\\
$\tau$&The distance threshold\\
$\gamma$&The time threshold\\
$\imath$&The headway threshold\\
$W_{e_k}$&The warning messages set at time ${e_k}$\\
$w_{v_i}^{e_k}$&The warning message indicator of vehicle ${v_i}$\\
&at time ${e_k}$\\
\hline
\hline
\end{tabular}
\label{table_notations}
\end{table}

\subsection{Fitting of Application-layer Transmission Latency}\label{delay fitting}
 
\begin{figure}
\centering
  \includegraphics[width=3.0in]{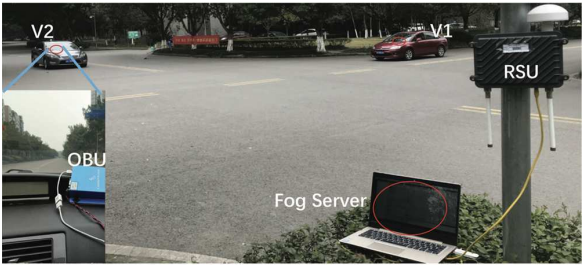}
  \caption{Real-world field testing scenario at an intersection without traffic lights}
  \label{fig_testing_scenario}
\end{figure}

We analyze the application-layer transmission latency of DSRC based on real-world field-testing data. 
In Fig. \ref{fig_testing_scenario}, a field testing is carried out on a road intersection without traffic lights.
Two vehicles equipped DSRC interfaces are able to communicate with RSU via V2I communication.
In particular, we utilize Cohda Wireless MK5 RSU and OBU as DSRC communication interfaces. 
An RSU is installed at the intersection, which is connected with a notebook as a computation unit of fog node. 
Vehicles $V_1$ and $V_2$ are approaching the intersection, and they are sending vehicle status including GPS coordinates, velocity, acceleration, heading direction and timestamp via V2I communication periodically.
Fog node receives the packet from vehicles and obtains the packet transmission latency of application-layer of V2I communication. 

In the following, we model the data packet transmission latency model in real vehicular communication environments with DSRC.
{\color{blue}
Based on the observation of transmission latency obtained by realistic field testing, we find that the Stable distribution that provides well-established model for non-Gaussian processes is suitable to fit the communication delay. (The fitting result shows it is convincing to fit the latency using the Stable distribution.)
}
We fit the observed V2I communication delays with the Stable distribution, which can be described by characteristics function as follows \cite{samoradnitsky2017stable},
\begin{equation}
\resizebox{.9\hsize}{!}{
$E \exp (i t X)\!=\!\left\{\begin{array}{l}{\exp \left\{-\sigma^{\alpha}|t|^{\alpha}[1-i \beta \tan (\alpha \pi / 2) \operatorname{sgn}(t)]+i \mu t\right\} \text { for } \alpha \neq 1} \\ {\exp \{-\sigma|t[1+i \beta(2 / \pi) \operatorname{sgn}(t) \ln (|t|)]+i \mu t\} \text { for } \alpha=1}\end{array},\right.$}
\label{equation_stable_distribution}
\end{equation}
\noindent
where $X$ is a random variable and follows the Stable distribution $X \sim {S(\alpha, \beta, \mu, \sigma)}$, 
and
\begin{equation}
\begin{array}{l}{0<\alpha \leq 2},{-1 \leq \beta \leq 1},{\mu \in \mathbb{R}},{\sigma>0},\end{array}
\end{equation}
$\alpha$ is an index of stability or characteristic exponent, when $\alpha = 2$, the relevant Stable distribution is Gaussian. 
$\beta$ is a skewness parameter, when $\beta = 0$, the relevant Stable distribution is symmetric about the center $\mu$, which is a location parameter. 
If $\alpha \neq 1$, the cases $\beta > 0$ and $\beta < 0$ correspond to left-skewness and right-skewness, respectively.
$\sigma$ is a scale parameter, and behaves like the variance.
The characteristic function $\phi(t) = E \exp (i t X)$ completely determines the behavior and properties of the probability distribution of the random variable $X$, where $t$ is a real number, $i$ is the imaginary unit and $E$ is the expected value. 
And $\operatorname{sgn}(t)$ is a sign function as defined by
\begin{equation}
\operatorname{sgn}(t)=\left\{\begin{array}{l}{1 \text { for } t>0} \\ {0 \text { for } t=0} \\ {-1 \text { for } t<0}\end{array}\right.
\end{equation}
We introduce the parameters estimation of the Stable distribution as follows.

We employ a regression-type estimation to estimate four parameters of the Stable distribution.
First, given observation data as a random simple $x_1, x_2, \cdots, x_n$ of size $n$, a simple characteristic function $\hat{\phi}(t)$ is defined by
\begin{equation}
\begin{aligned}
\hat{\phi}(t)&\!=\!\frac{1}{n} \sum_{j=1}^{n} \exp \left(i t x_{j}\right)\\
&\!=\!\frac{1}{n} \sum_{j=1}^{n}\left[\cos \left(t x_{j}\right)+\sin \left(t x_{j}\right) i\right]\\
&\!=\!\frac{1}{n} \sum_{j=1}^{n} \cos \left(t x_{j}\right)+i \frac{1}{n} \sum_{j=1}^{n} \sin \left(t x_{j}\right)
\end{aligned}
\end{equation}
when $\alpha \neq 1$, we denote:
\begin{equation}
\resizebox{.9\hsize}{!}{$\begin{aligned}
\phi(t) 
&\!=\!E \exp (i t X) \\ 
&\!=\!\exp \left\{-\sigma^{\alpha}|t|^{\alpha}[1\!-\!i \beta \tan (\alpha \pi / 2) \operatorname{sgn}(t)]+i \mu t\right\} \\ 
&\!=\!\exp \left\{-\sigma^{\alpha}|t|^{\alpha}\!+\!\left[\mu t\!+\!\sigma^{\alpha}|t|^{\alpha} \beta \tan (\alpha \pi / 2) \operatorname{sgn}(t)\right] i\right\} \\ 
&\!=\!\exp \left(-\sigma^{\alpha}|t|^{\alpha}\right) \exp \left[\left(\mu t\!+\!\sigma^{\alpha}|t|^{\alpha} \beta \tan (\alpha \pi / 2) \operatorname{sgn}(t)\right] i\right\} \\ 
&\!=\!\exp \left(-\sigma^{\alpha}|t|^{\alpha}\right) \cos \left[\mu t\!+\!\sigma^{\alpha}|t|^{\alpha} \beta \tan (\alpha \pi / 2) \operatorname{sgn}(t)\right] \\ 
&\!+\!\exp \left(-\sigma^{\alpha}|t|^{\alpha}\right) \sin \left[\mu t\!+\!\sigma^{\alpha}|t|^{\alpha} \beta \tan (\alpha \pi / 2) \operatorname{sgn}(t)\right] i 
\end{aligned}$}
\label{Equation_phi_t}
\end{equation}

The distribution is assumed symmetric about the center $0$ ($\beta = 0, \mu = 0$), then we can easily obtain:
\begin{equation}
-\ln |\phi(t)|^{2}=2 \sigma^{\alpha}|t|^{\alpha}
\end{equation}
\begin{equation}
\begin{aligned} \ln \left(-\ln | \phi(t)^{2}\right)
&= \ln \left(2 \sigma^{\alpha}|t|^{\alpha}\right) \\ 
&=\ln \left(2 \sigma^{\alpha}\right)+\alpha \ln (|t|) \end{aligned}
\end{equation}

We estimate $\alpha$ and $\sigma$ by regressing $y_{k}=\alpha \omega_{k}+b$, where ${b=\ln \left(2 \sigma^{\alpha}\right)}$ and ${\omega_{k}=\ln \left(\left|t_{k}\right|\right)}$.
We denote $f\left(t_{k}\right)=\ln \left(-\ln \left|\phi\left(t_{k}\right)\right|^{2}\right)$ and use linear regression to address estimation problem by minimizing the mean-square error.
\begin{equation}
\resizebox{.9\hsize}{!}{$
\begin{aligned}\left(\alpha^{*}, b^{*}\right) &=\underset{(\alpha, b)}{\arg \min } \sum_{k=1}^{K}\left(f\left(t_{k}\right)-y_{k}\right)^{2} \\ &=\underset{(\alpha, b)}{\arg \min } \sum_{k=1}^{K}\left[\ln \left(-\ln \left|\phi\left(t_{k}\right)\right|^{2}\right)-\left(\alpha \omega_{k}+b\right)\right]^{2}  \end{aligned}$}
\end{equation}

Then, we use the least square method to get estimation value $\hat{\alpha}$ and $\hat{\sigma}$.
We denote $E_{(\alpha, b)}=\sum_{k=1}^{K}\left(f\left(t_{k}\right)-y_{k}\right)^{2}$ and the estimated parameters are obtained by resolving the following equations:
\begin{equation}
\resizebox{.9\hsize}{!}{$
\left\{\begin{aligned} \frac{\partial E_{(\alpha, b)}}{\partial \alpha} &=2 \left(\alpha \sum_{k=1}^{K} \omega_{k}^{2}-\sum_{k=1}^{K}\left(f\left(t_{k}\right)-b\right) \omega_{k}\right) &=0 \\ 
\frac{\partial E_{(\alpha, b)}}{\partial b} &=2\left(K b-\sum_{k=1}^{K}\left(f\left(t_{k}\right)-\alpha \omega_{k}\right)\right) &=0 \end{aligned}\right.$}
\end{equation}

The estimated parameters $\hat{\alpha}$ and $\hat{\sigma}$ are represented by:
\begin{equation}
\label{equation_alpha_sigma}
\left\{\begin{array}{l}{\hat{\alpha}={ \sum_{k=1}^{K} f\left(t_{k}\right)\left(\omega_{k}-\bar{\omega}\right)}/{\sum_{k=1}^{K} \omega_{k}^{2}-\frac{1}{K}\left(\sum_{k=1}^{K} \omega_{k}\right)^{2}} } 
\\ {\hat{\sigma}=\sqrt[\hat{\alpha}]{ (\exp \hat{b}) / 2}}
\end{array}\right.
\end{equation}
where
${\hat{b}=\frac{1}{K} \sum_{k=1}^{K}\left(f\left(t_{k}\right)-\hat{\alpha} \omega_{k}\right)}$
and $\bar{\omega}=\frac{1}{K} \sum_{k=1}^{K} \omega_{k}$.

It is easy to see that the real and imaginary parts of $\phi(t)$, $\operatorname{Re} \phi(t)$ and $\operatorname{Im} \phi(t)$, are implied from Equation. \ref{Equation_phi_t}.
\begin{equation}
\resizebox{.9\hsize}{!}{$\operatorname{Re} \phi(t)=\exp \left(-\sigma^{\alpha}|t|^{\alpha}\right) \cos \left[\mu t+\sigma^{\alpha}|t|^{\alpha} \beta \tan (\alpha \pi / 2) \operatorname{sgn}(t)\right]$}
\end{equation}
\begin{equation}
\resizebox{.9\hsize}{!}{$\operatorname{Im} \phi(t)=\exp \left(-\sigma^{\alpha}|t|^{\alpha}\right) \sin \left[\mu t+\sigma^{\alpha}|t|^{\alpha} \beta \tan (\alpha \pi / 2) \operatorname{sgn}(t)\right]$}
\end{equation}
The last two equations lead to
\begin{equation}
\resizebox{.9\hsize}{!}{$
\arctan \left(\frac{\operatorname{Im} \phi(t)}{\operatorname{Re} \phi(t)}\right)=\mu t+\sigma^{\alpha}|t|^{\alpha} \beta \tan (\alpha \pi / 2) \operatorname{sgn}(t).$}
\end{equation}

As the estimated parameters $\hat{\alpha}$ and $\hat{\sigma}$ are obtained according to Equation. \ref{equation_alpha_sigma}, we estimate another two parameters $\beta$ and $\mu$ by regressing $q_{l}=\mu+c d_{l}$, where ${d_{l}=\operatorname{sgn}\left(t_{l}\right)\left|t_{l}\right|^{\alpha-1}}$ and ${c=\sigma^{\alpha} \beta \tan (\alpha \pi / 2)}$.
And we denote $\varphi\left(t_{l}\right)=\frac{1}{t_{l}} \arctan \left(\frac{\operatorname{Im} \phi\left(t_{l}\right)}{\operatorname{Re} \phi\left(t_{l}\right)}\right)$ and minimize the mean-square error between $\varphi\left(t_{l}\right)$ and $q_{l}$.
\begin{equation}
\resizebox{.9\hsize}{!}{$
\begin{aligned}\left(c^{*}, \mu^{*}\right) &=\underset{(c, \mu)}{\arg \min } \sum_{l=1}^{L}\left(\varphi\left(t_{l}\right)-q_{l}\right)^{2} \\ &=\underset{(c, \mu)}{\arg \min } \sum_{l=1}^{L}\left(\frac{1}{t_{l}} \arctan \left(\frac{\operatorname{Im} \phi\left(t_{l}\right)}{\operatorname{Re} \phi\left(t_{l}\right)}\right)-\mu-c d_{l}\right)^{2}  \end{aligned}$}
\end{equation}

We denote $E_{(c, \mu)}=\sum_{l=1}^{L}\left(\varphi\left(t_{l}\right)-q_{l}\right)^{2}$, and the solution can obtained by resolving the follows:
\begin{equation}
\resizebox{.9\hsize}{!}{$
\left\{\begin{aligned} \frac{\partial E_{(e, \mu)}}{\partial c}&=2\left(c \sum_{l=1}^{L} d_{l}^{2}-\sum_{l=1}^{L}\left(\varphi\left(t_{l}\right)-\mu\right) \cdot d_{l}\right)&=0 
\\ \frac{\partial E_{(c, \mu)}}{\partial \mu}&=2\left(L \mu-\sum_{l=1}^{L}\left(\varphi\left(t_{l}\right)-c d_{l}\right)\right)&=0 \end{aligned}\right.$}
\end{equation}

The estimated parameters $\hat{\beta}$ and $\hat{\mu}$ are obtained by:
\begin{equation}
\label{Equation_beta_mu}
\left\{\begin{aligned} \hat{\beta}=& \frac{\hat{c}}{\hat{\sigma}^{\hat{\alpha}} \tan (\hat{\alpha} \pi / 2)}
\\ \hat{\mu}=& \frac{1}{L} \sum_{l=1}^{L}\left(\varphi\left(t_{l}\right)-\hat{c} d_{l}\right) \end{aligned}\right.
\end{equation}
where 
\begin{equation}
\hat{c}=\frac{\sum_{l=1}^{L} \varphi\left(t_{l}\right)\left(d_{l}-\bar{d}\right)}{\sum_{l=1}^{L} d_{l}^{2}-\frac{1}{L}\left(\sum_{l=1}^{L} d_{l}\right)^{2}}, 
\end{equation}
and $\bar{d}=\frac{1}{L} \sum_{l=1}^{L} d_{l}$.

\begin{figure}
\centering
  \includegraphics[width=3.0in]{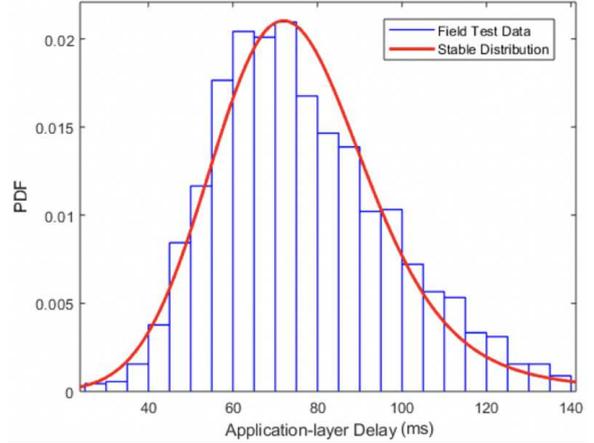}
  \caption{Probability density function (PDF) of application-layer transmission delay of DSRC, where $\alpha  = 1.77395, \beta  = 1, \mu  = 72.7343, \sigma  = 13.3685$}
  \label{fig_3}
\end{figure}

With equations \ref{equation_alpha_sigma} and \ref{Equation_beta_mu}, and given the observed data set $X^{(0)}$, we obtain the four estimated parameter of the Stable distribution.
First, the given observed data set is $X^{(0)} = (x_1^{(0)}, x_2^{(0)}, \cdots, x_n^{(0)})$.
In the $pth$ iteration, we standardize the data by:
\begin{equation}
x_{j}^{(p)}=\left(x_{j}^{(p-1)}-\mu_{p-1}\right) / \sigma_{p-1}, p = 1,2,\cdots
\end{equation}
where ${\sigma_{0}=\left(x_{.72}-x_{.28}\right) / 1.654}$ and $\mu_{0}=25 \%$ truncated mean of origin data, $x_{f}$ is $f$ sample quantile \cite{fama1971parameter}.
We choose optimum $K^{(p)}$ point $t_{k}=\pi k / 25, k=1,2,\cdots,K^{(p)}$ \cite{koutrouvelis1980regression} to estimate $\hat{\alpha}^{(p)}$ and $\hat{\sigma}^{(p)}$ in Equation. \ref{equation_alpha_sigma}.

\begin{equation}
{\hat{\alpha}^{(p)}=\frac{ \sum_{k=1}^{K^{(p)}} f\left(t_{k}\right)\left(\omega_{k}-\bar{\omega}\right)}{\sum_{k=1}^{K^{(p)}} \omega_{k}^{2}-\frac{1}{K}\left(\sum_{k=1}^{K^{(p)}} \omega_{k}\right)^{2}} } 
\end{equation}
\begin{equation}
{\hat{\sigma}^{(p)}=\sqrt[\hat{\alpha}^{(p)}]{ (\exp \hat{b}^{(p)}) / 2}}
\end{equation}
where $f\left(t_{k}\right) = \ln \left(-\ln \left|\frac{1}{n} \sum_{j=1}^{n} \exp \left(i t_{k} x_{j}^{(p)}\right)\right|^{2}\right)$.

And these estimated parameters are in turns to obtains another two estimated parameters $\hat{\beta}^{(p)}$ and $\hat{\mu}^{(p)}$ with optimum $L^{p}$ points $t_{l}=\pi l / 25, l=1,2,\cdots,L^{(p)}$ \cite{koutrouvelis1980regression} in Equation. \ref{Equation_beta_mu}.
\begin{equation}
\hat{\beta}^{(p)}= \frac{\hat{c}^{(p)}}{\bar{\sigma}^{\bar{\alpha}} \tan (\bar{\alpha} \pi / 2)}
\end{equation}
\begin{equation}
\hat{\mu}^{(p)}= \frac{1}{L^{(p)}} \sum_{l=1}^{L^{(p)}}\left(\varphi\left(t_{l}\right)-\hat{c}^{(p)} d_{l}\right)
\end{equation}
where $\bar{\alpha} =  {\hat{\alpha}^{(p)}}$ , $\bar{\sigma} =  {\hat{\sigma}^{(p)}}$ and 
\begin{equation}
\varphi\left(t_{l}\right) = \frac{1}{t_{l}} \arctan \left(\frac{ \sum_{j=1}^{n} \sin \left(t_l x_{j}^{(p)}\right)}{\sum_{j=1}^{n} \cos \left(t_l x_{j}^{(p)}\right)}\right).
\end{equation}

After finite iteration, we obtain the four estimated parameters which are satisfy required accuracy.
{\color{blue}
We estimate the Stable distribution model using 1804 data packet transmissions latency obtained by real-world field testing.
Fig. \ref{fig_3} shows the probability density function (PDF) of the application-layer delay.
The distribution of application-layer delay is almost symmetric ($\alpha  = 1.77395$) about the Mean ($\mu  = 72.7343$).
We are 95 percent confident that the true Mean lies within the interval $71.9384$ and $73.5301$.
As can be seen, the resulting distribution is characterized by the left-skewness ($\beta  = 1$) and the average of the not small squared differences from the Mean ($\sigma  = 13.3685$).
}

\subsection{Packet Loss Detection Method}\label{packet loss detection}
We propose a packet loss detection method in fog nodes based on the historical information including data transmission frequency and locations of vehicles.
The general idea is described as follows. 
Each fog node obtains the ID set of vehicles, which are supposed to report their status according to uploading frequency.
If the fog node does not receive the packets which are expected.
There are two possible cases.
First, the vehicle is out of DSRC communication range, thus, the packet cannot be delivered successfully.
Second, the vehicle is within the DSRC communication range, however, the packet is lost in the transmission.
For those packets, fog node figures out whether the vehicle is out of communication range according its historical position, if not, fog node supposes that the packet is lost.

Without loss of generality, we consider that the system consists of one fog node and a number of vehicles. 
Note that it can be straightforwardly extended to the scenario of multiple fog nodes. 
In the concerned scenario, we use set $E=(e_1, e_2, \cdots, e_{|E|})$ to denote the time slots, where $e_{k+1} - e_{k} = \frac{1}{\xi}$, $k=1,2,\cdots,|E|$, and $|E|$ is the number of the time slot set.
We use set $V=(v_{1}, v_{2}, \cdots, v_{|V|})$ to denote vehicles, where $|V|$ is the number of vehicles.
At time $e_r$, the position, velocity, acceleration, heading direction of vehicle $v_{i}$ are denoted by $l_{v_i}^{e_r}$, $s_{v_i}^{e_r}$, $a_{v_i}^{e_r}$ and $d_{v_i}^{e_r}$, respectively.
The location of the fog node is denoted by ${l}_{f}$.
And the communication range of DSRC is denoted by $R$.
he distance between vehicle ${v_{i}}$ and the fog node at time $e_r$ is denoted by $dis({l_{v_{i}}^{e_r}}, {l_{f}})$.
If $dis({l_{v_{i}}^{e_r}}, {l_{f}}) \leq R$, vehicle ${v_{i}}$ can communicate with the fog node.

The fog node receives a number of packets denoted by $M_{e_{k}}={(m_1, m_2, \cdots, m_{|{M}_{e_{k}}|})}$ at time $e_k$, $e_k \in E$, where $m_{q}=({{{l}_{v_{i}}^{e_{r}}}, {{s}_{v_{i}}^{e_{r}}}, {{a}_{v_{i}}^{e_{r}}}, {{d}_{v_{i}}^{e_{r}}}})$, $m_{q} \in {M}_{e_{k}}$ and $|{M}_{e_{k}}|$ is the number of the packets set.
Meanwhile, fog node records the received packets of each time slot.
At time $e_{k}$, we use set ${H_{e_{k}}} = ({M_{e_{1}}},{M_{e_{2}}}, \cdots, {M_{e_{k-1}}})$ to denote the historical records.
The packet loss detection method is departed into two steps.

$Record):$ Fog node maintains a vehicles ID set ${ID}_{e_k}$, which records ID of all vehicles under the coverage of the fog node at time $e_k$.
${{ID}_{e_k}}$ initialize as ${{ID}_{e_k}}={{ID}_{e_{k-1}}}$.
{\color{blue}
When the fog node receives a number of packets $M_{e_k}$ at time slot $e_k$.
For $m_{q}=({{{l}_{v_{i}}^{e_{r}}}, {{s}_{v_{i}}^{e_{r}}}, {{a}_{v_{i}}^{e_{r}}}, {{d}_{v_{i}}^{e_{r}}}})$, ${m_{q}} \in M_{e_k}$, if the fog node receives the packet of vehicle $v_{i}$ at the first time, ${v_{i}} \notin {ID}_{e_{k}}$, then adds ${{v_{i}}}$ into ${ID}_{e_{k}}$, ${{ID}_{e_{k}}} = {ID}_{e_{k}} \cup \{{v_{i}}\}$.
Fog node searches $M_{e_{k}}$ and adds all vehicle ID into the set ${ID}_{M_{e_{k}}}$.}

$Detection):$ 
For the vehicle ${v_{i}}$, ${v_{i}} \in {ID}_{e_{k}} \cup {{ID}_{M_{e_{k}}}}$, there are two cases:
a) ${v_{i}} \in {{ID}_{e_{k}}} \setminus {{ID}_{M_{e_{k}}}}$, vehicle $v_i$ can communicate with fog node, however, fog node does not receive the packet from it;
b) ${v_{i}} \in {{ID}_{e_{k}}} \cap {{ID}_{M_{e_{k}}}}$, vehicle $v_i$ can communicate with fog node, and fog node receives its packet.
Therefore, for case a, fog node searches ${H_{e_{k}}}$ to get its laster new position ${{l}_{v_{i}}^{e_{r}}}$.
Fog node use a distance threshold $\tau$ and a time threshold $\gamma$ to detect whether the vehicle is out of communication range.
If the vehicle is within the communication range, fog node detects whether the packet is lost in transmission.
If the distance between vehicle ${v_{i}}$ and fog node $dis({l_{v_{i}}^{e_r}}, {l_{f}}) \geq R - \tau$, that means vehicle ${v_{i}}$ is leaving the communication range.
So that, fog node remove ${v_{i}}$ from ${{ID}_{e_{k}}}$, ${{ID}_{e_{k}}}={{ID}_{e_{k}}} \setminus \{{v_{i}}\}$.
If $dis({l_{v_{i}}^{e_r}}, {l_{f}}) < R - \tau$, that means vehicle $v_i$ can communicate with fog node, however, fog node does not receive its packet.
The estimated received time is ${e_{r}}+\frac{1}{\xi}$, if ${e_{k}} - {e_{r}} > \frac{1}{\xi} + \gamma$, fog node assume the packet is lost.
Otherwise, vehicle $v_{i}$ does not send the packet yet, or the packet is delay due to the wireless communication.

\subsection{Procedures of TCCW Algorithm}\label{tccw algorithm}

The general procedures of TCCW are described as follows.
First, we estimate the transmission latency of each data packet and update their real-time status according to their velocity and acceleration.
Second, we detect the lost packets and update their status using packets records in the fog node.
Third, we calibrate vehicle trajectories using simulated transmission latency.
Then, we predict the future trajectories based on calibrated trajectories.
Finally, we predict the potential collisions by calculating the headway of each pair of vehicles and using the headway threshold.
At time ${e_{k}}$, vehicle ID set ${ID}_{e_{k}}$, received packets $M_{e_{k}}$, packets records $H_{e_{k}}$ are known as the input of the algorithm.
The output of TCCW is the warning message set $W_{e_{k}}$.
The detailed procedures of TCCW are presented as follows and the pseudo-code is shown in Algorithm. \ref{algorithm1}.

\begin{algorithm}[htb]
\caption{TCCW}\label{algorithm1}
\begin{algorithmic}[1]
\REQUIRE ~~\\
$M_{e_{k}}={(m_1, m_2, \cdots, m_{|{M}_{e_{k}}|})}$, the received packets set. \\
${H_{e_{k}}} = ({M_{{e_1}}},{M_{{e_2}}}, \cdots, {M_{{e_{k-1}}}})$, the historical packets set.\\
${{ID}_{e_k}}$, vehicle ID set at time ${e_k}$. 
\ENSURE ${W_{e_{k}}}=\{ {w_{e_k}^{v_{1}}},{w_{e_k}^{v_{2}}}, \cdots ,{w_{e_k}^{v_{|V|}}}\}$~~\\
\STATE ${{ID}_{e_k}}={{ID}_{e_{k-1}}}$
\STATE ${{ID}_{M_{e_{k}}}} = \{\}$
\FOR{${{m}_{p}} \in {M_{e_{k}}}$}
\STATE ${{ID}_{M_{e_{k}}}}={{ID}_{M_{e_{k}}}} \cup \{{v_i}\}$
\IF{${{v_i}} \notin {{ID}_{e_k}}$}
\STATE ${{ID}_{e_k}} = {{ID}_{e_k}} \cup \{{v_i}\}$
\ENDIF
\ENDFOR
\FOR{${{v_i}} \in {{ID}_{e_k}} \cup {{ID}_{M_{e_{k}}}}$}
\IF{${v_{i}} \in {{ID}_{e_{k}}} \setminus {{ID}_{M_{e_{k}}}}$}
\STATE Search ${H_{e_{k}}}$ and get latest new packet $m_q$
\IF{$dis({{l}_{v_{i}}^{e_{r}}}, {l_{f}}) \geq r - \tau$}
\STATE ${{ID}_{e_k}} = {{ID}_{e_k}} \setminus \{{v_i}\}$
\ELSE
\IF{${e_{k}} - {e_{r}} > \frac{1}{\xi} + \gamma$}
\STATE ${M_{e_{k}}} = {M_{e_{k}}} \cup \{ {{m}_{q}}\}$
\ENDIF
\ENDIF
\ENDIF
\ENDFOR
\FOR{${{m}_{v_{i}}^{e_{r_p}}} \in {M_{e_{b}}}$}
\STATE ${e_{ts}} = {e_k} - {e_{r}} + {e_t^{v_i}}$
\STATE Update location of vehicle using Eq.\ref{equation_update_location}
\ENDFOR
\FOR{${{m}_{p}} \in {M_{e_{k}}}$}
\REPEAT
\STATE ${e_{ts}} = {e_{ts}} + \frac{1}{\xi}$
\STATE Calculate ${{l}_{v_i}^{e_{ts}}}$ using Eq.\ref{equation_update_location}
\STATE ${{Tra}_{v_i}} = {{Tra}_{v_i}} \cup \{ {{l}_{v_i}^{e_k}} \}$
\UNTIL{${e_{ts}} > {{e_k} + {e_{pre}}}$}
\STATE ${Tra} = {Tra} \cup \{{{Tra}_{v_i}}\}$
\ENDFOR
\FOR{${{Tra}_{v_i}} \in {Tra}$ and ${{Tra}_{v_j}} \in {Tra} \setminus \{ {{Tra}_{v_i}}\}$}
\FOR{${l_{v_{i}}^{e_{u}}} \in {{Tra}_{v_i}}$ and ${l_{v_{j}}^{e_{v}}} \in {{Tra}_{v_j}}$}
\IF{$dis({l_{v_{i}}^{e_{u}}}, {l_{v_j}^{e_{v}}}) < d_{col}$}
\STATE ${h}_{ij} = |e_{u} - e_{v}|$
\IF{${h}_{ij} < \imath$}
\STATE ${w_{v_i}^{e_k}} = {w_{v_j}^{e_k}} = 1$
\ENDIF
\ENDIF
\ENDFOR
\ENDFOR
\end{algorithmic}
\end{algorithm}

$1)\ Updating\ the\ vehicle\ set:$
The purpose of updating the vehicle set is to maintain the vehicle ID set ${{ID}_{e_{k}}}$.
We initialize vehicle ID set ${{ID}_{e_{k}}}$ at time ${e_k}$ and the ID set ${{ID}_{M_{e_k}}}$ of received packets ${M_{e_k}}$.
${{ID}_{M_{e_k}}}$ contains every vehicle ID in received packets, and if the vehicle ID is not contained by ${{ID}_{e_{k}}}$, then fog node add the vehicle ID into ${{ID}_{e_{k}}}$.
The detailed procedure of updating the vehicle set is shown in lines 1-8 in Algorithm. \ref{algorithm1}.

$2)\ Detecting\ packet\ loss:$
Fog node detects the packet loss in transmission using the method introduced in Section. \ref{packet loss detection}.
For those vehicles that fog node do not received from them, fog node searches packets records ${H}_{e_a}$ to get last new vehicle state information, and add it into ${M_{e_b}}$.
The detailed procedure of detecting packet loss is shown in lines 9-20 in Algorithm. \ref{algorithm1}.

$3)\ Calibrating\ vehicle\ trajectories:$
For the packet ${{m}_{p}}$, ${{m}_{p}} \in M_{e_{k}}$, fog node estimates the transmission latency $e_{t}^{v_{i}}$ of the packet by generating a random number using the Stable distribution.
Fog node estimates the packet send time ${\hat e_{c}} = {e_{r}} - e_{t}^{v_{i}}$. 
The interval between time ${e_{k}}$ and packet send time ${\hat e_{c}}$ is ${e_{ts}} = {e_{k}} - {e_{r}} + e_{t}^{v_{i}}$.
Fog node updates location information of vehicle ${v_{i}}$ as follows:
\begin{equation}
\left\{\begin{array}{ll}{{l_x}_{v_{i}}^{e_{k}}} = {{l_x}_{v_{i}}^{e_{r}}} + {e_{ts}}{{s_x}_{v_{i}}^{e_{r}}} + \frac{1}{2}{e_{ts}^2}{{a_x}_{v_{i}}^{e_{r_p}}} \\ 
{{l_y}_{v_{i}}^{e_{k}}} = {{l_y}_{v_{i}}^{e_{r}}} + {e_{ts}}{{s_y}_{v_{i}}^{e_{r}}} + \frac{1}{2}{e_{ts}^2}{{a_y}_{v_{i}}^{e_{r_p}}}\end{array}\right.
\label{equation_update_location}
\end{equation}
\noindent
where ${l_x}_{v_{i}}^{e_{k}}$, ${l_y}_{v_{i}}^{e_{k}}$, ${s_x}_{v_{i}}^{e_{k}}$, ${s_y}_{v_{i}}^{e_{k}}$, ${a_x}_{v_{i}}^{e_{k}}$, ${a_y}_{v_{i}}^{e_{k}}$ are positions, speeds, accelerations of vehicle $v_i$ in X and Y coordinate respectively.
Lines 21-24 in Algorithm. \ref{algorithm1} show the detailed procedure of calibrating vehicle trajectories.

$4)\ Predicting\ future\ trajectories:$
For vehicle $v_i$, fog node predicts the future trajectories in time slot $({e_{k}},{e_{k}} + {e_{pre}})$, where ${e_{pre}}$ is fog node prediction time.
Fog node calculates the vehicle position in every $\frac{1}{\xi}$ second, and appends the position into the trajectory set ${Tra}_{v_i}$ of vehicle $v_i$.
The detailed procedure of predicting future vehicle trajectories is shown in lines 25-32 in Algorithm. \ref{algorithm1}.

$5)\ Predicting\ potential\ collisions:$
Fog node use a set of warning message ${W_{e_{k}}}=\{ {w_{v_1}^{e_{k}}},{w_{v_2}^{e_{k}}}, \cdots ,{w_{v_{|V|}}^{e_k}}\}$, ${w_{v_i}^{e_k}}$ is a binary variable to indicate whether vehicle $v_i$ has the potential collision.
For ${l_{v_{i}}^{e_{u}}} \in {{Tra}_{v_i}}$ and ${l_{v_{j}}^{e_{v}}} \in {{Tra}_{v_j}}$, fog node calculates the distance of the two vehicles $dis({l_{v_{i}}^{e_{u}}}, {l_{v_j}^{e_{v}}})$.
We assume that vehicle $v_i$ and $v_j$ pass the same point if the distance $dis({l_{v_{i}}^{e_{u}}}, {l_{v_j}^{e_{v}}}) < d_{col}$.
The elapsed time between the front of the lead vehicle passing a point on the roadway and the front of the following vehicle passing the same point is defined as the headway \cite{vogel2003comparison}.
Fog node calculates the headway ${h}_{ij} = |e_{u} - e_{v}|$, and if ${h}_{ij} < \imath$, where $\imath$ is the headway threshold, warning message will be triggered, ${w_{v_i}^{e_k}} = {w_{v_i}^{e_k}} = 1$.
Lines 32-42 in Algorithm. \ref{algorithm1} show the detailed procedure of predicting potential collisions.

\section{Performance Evaluation}\label{Performance Evaluation}
\subsection{Experiment Setup}

\begin{figure}
\centering
  \includegraphics[width=3.0in]{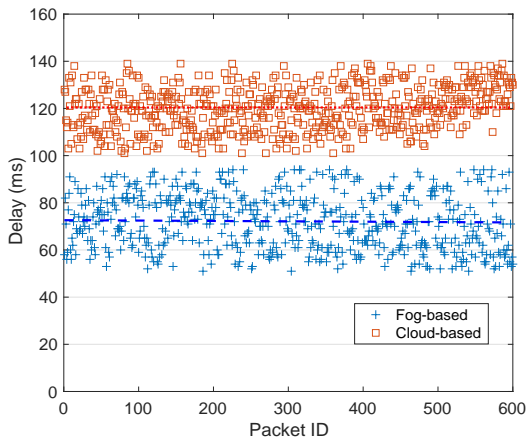}
  \caption{Transmission delay under different system architectures where average delay of packets transmission in cloud and fog computing is 120${ms}$ and 77${ms}$, respectively}
  \label{fig_4}
\end{figure} 

To further demonstrate the performance of our algorithm, we carry out another real-world field testing, and obtain 600 packet transmission delay via two different architectures, both fog computing and cloud computing architecture. 
In the field testing experiment, when a vehicle is approaching the road intersection, it uploads its status including GPS coordinates, velocity, acceleration, heading direction to fog node.
In detail, the vehicle uploading frequency is 1 HZ, in another word, vehicles send packets in every one second.
600 packets, which are transmitted to fog node via DSRC and cloud node via 4G-LTE respectively, are collected and analyzed. 
Fig. \ref{fig_4} shows that the average delay of packets transmission in cloud computing architecture is 120${ms}$, otherwise, the average delay of the fog computing architecture is 77${ms}$.
It is obvious that packet transmission latency in the fog computing based architecture is lower than that in the cloud computing based.
Undoubtedly, the results conclusively demonstrate the superiority of vehicular fog computing enabled collision warning system for supporting low latency and safety-critical services.
The real-world communication latency obtained by a field-testing experiment is utilized as transmission delay of each packets in the simulation experiment. 

\begin{figure*}
\centering
    \subfigure[Fog node located in $(10422.0, 12465.3)$]{
    \begin{minipage}{3.0in}
    \centering
    \label{fig_experiment_one}
    \includegraphics[scale=0.25]{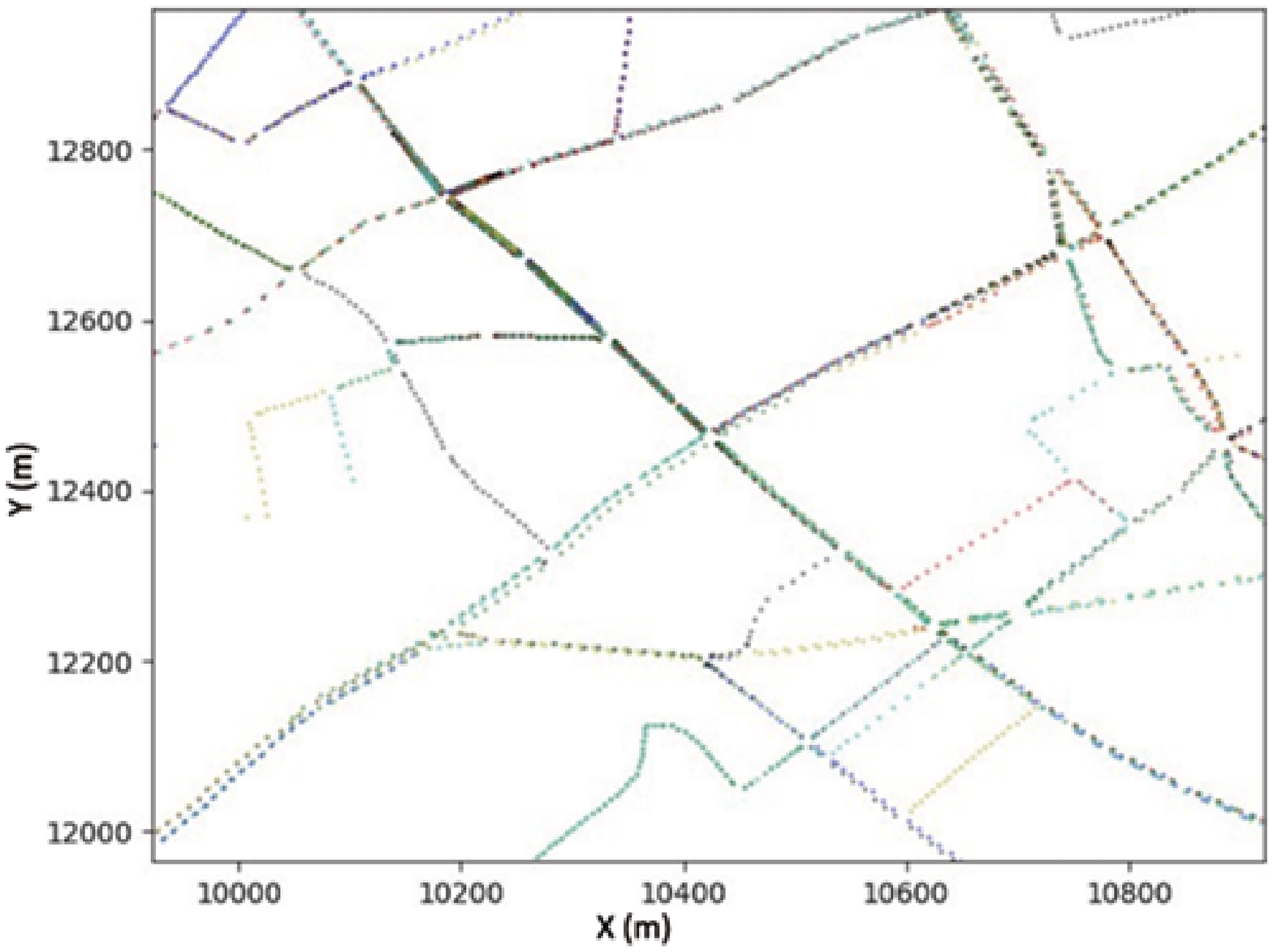}
    \end{minipage}}
    \subfigure[Fog node located in $(6097.1, 14870.0)$]{
    \begin{minipage}{3.0in}
    \centering
    \label{fig_experiment_two}
    \includegraphics[scale=0.25]{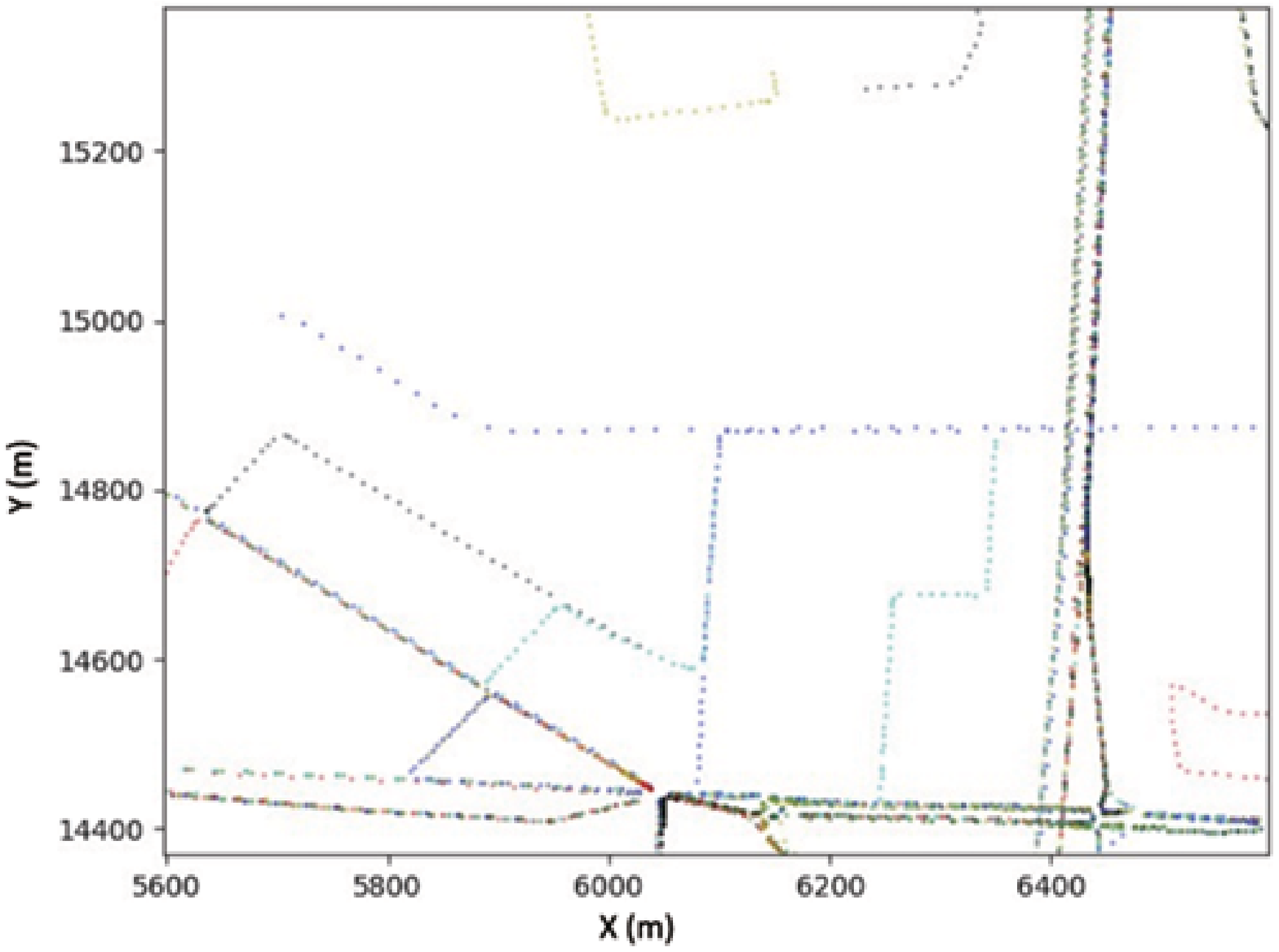}
    \end{minipage}}
\caption{Vehicular trajectory in two road intersections without traffic lights in 7 PM and 6 PM, respectively}
\label{fig_experiment}
\end{figure*}

First, we employ a vehicle mobility dataset of real taxi trajectory in the experiment.
The dataset is collected on an area of about 400 square kilometers from Cologne, Germany for a period of 24 hours in a typical working day \cite{uppoor2013}.
There are more than 1.2 million individual vehicle trajectories, which contains 3.5 billion points of vehicular positions.
The dataset contains vehicles positions in a broad range as a whole city, we choose some different road intersection to implement the system, therefore, the experiment scenarios have different traffic conditions. 
The features of five scenarios are listed in Table. \ref{table}.
We evaluate the traffic features with different scenarios in terms of vehicle number, average speed, and average acceleration.
Vehicle number is a variable to generally describe vehicular density of the scenario using a number of vehicles on the road or pass through in a period.
Average speed is the average speed of vehicles in the scenario.
As follows is specific data of vehicle number, average speed and average acceleration in five selected scenarios.
In scenarios 1, 2 and 3, fog node is located in a road intersection shown in Fig. \ref{fig_experiment_one}, and scenarios 1, 2, 3 start the experiment at  10 PM, 8 AM and 7 PM, respectively.
The fog node is located in center of Fig. \ref{fig_experiment_two} in scenario 4 and 5. 
It starts the experiment at 4 PM and 6 PM.
Fog node is installed in $(10422.0, 12465.3)$ of the first three scenarios and  in $(6097.1, 14870.0)$ of the last two scenarios.
X and Y coordinates represent location of vehicle with two-dimensional in meters.

\begin{table*}[b]
\centering
\caption{Different features in selected scenarios}
\label{table}
\setlength{\tabcolsep}{7mm}
\begin{tabular}{llllll}
\toprule
Scenario&1&2&3&4&5\\
\midrule
Vehicle Number& 54& 81& 106&	 85&	 114\\
Average Speed($km/h$)& 50.44& 46.58& 38.16& 69.19& 69.05\\
Average Acceleration($m/s^2$)& 0.203& 0.007& 0.075& 0.165& 0.060\\	
\bottomrule
\end{tabular}
\end{table*}

Apart from the five selected scenarios, we also implement the system under different headway and packet loss rate. 
Headway is a time slot between two vehicles passes through one point, but in the experiment, there is a headway threshold to determine whether the potential collision warning happened. 
If the headway threshold is small, the prediction collision number will less.  
The duration of each experiment is 100 seconds, and DSRC communication range is 500 meters. 
We conduct three sets of experiments, for each set of experiments, we change different parameters including scenario, headway and packet loss rate. 
We can find that performance of proposed system in several scenarios where traffic conditions are different. 

In specific, there are predicted collision warning message set ${W_{p}}$ and expected collision warning set ${W_{d}}$ in the experiment.
The number of expected warnings $\left| {W_{d}} \right|$ is the number of expected collision warning according to experimental setup, the number of predicted collision warning message $\left| {W_{p}} \right|$ is the number of potential collisions predicted by the collision warning system. 
Further, $\left| {{W_{d}} \cap {W_{p}}} \right|$ means the number of expected collision warnings which are actual predicted success, in another word, the number of successful predictions. 
$\left| {{W_{d}} - {W_{p}}} \right|$ is the number of expected warnings which should be triggered but not be predicted successful by collision warning system, in a word, it is failed predictions in ${W_{p}}$. 
In the same, $\left| {{W_{p}} - {W_{d}}} \right|$ is the number of wrong predictions, namely, it is potential collision predicted by collision warning system, however, it is not expected collision warning. 
We define precision and recall as follows:
\begin{equation}
precision = \frac{{\left| {{W_{d}} \cap {W_{p}}} \right|}}{{\left| {{W_{d}} \cap {W_{p}}} \right| + \left| {{W_{p}} - {W_{d}}} \right|}}
\label{equation_precision}
\end{equation}
\begin{equation}
recall = \frac{{\left| {{W_{d}} \cap {W_{p}}} \right|}}{{\left| {{W_{d}} \cap {W_{p}}} \right| + \left| {{W_{d}} - {W_{p}}} \right|}}
\label{equation_recall}
\end{equation}

\begin{figure*}
\centering
    \subfigure[Precision with different scenarios]{
    \begin{minipage}{3.0in}
    \centering
    \label{fig_experiment_scenario_one}
    \includegraphics[scale=0.40]{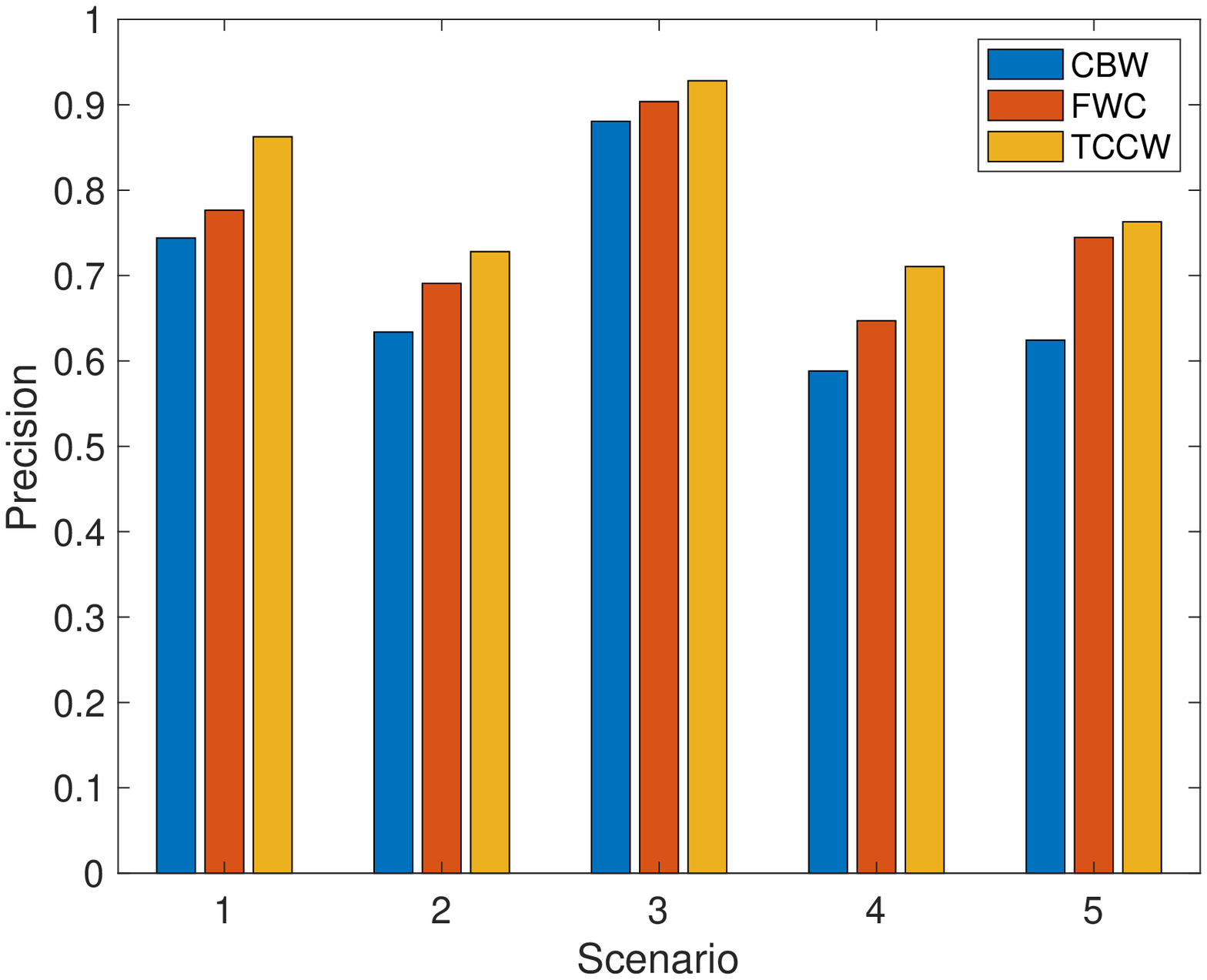}
    \end{minipage}}
    \subfigure[Recall with different scenarios]{
    \begin{minipage}{3.0in}
    \centering
    \label{fig_experiment_scenario_two}
    \includegraphics[scale=0.40]{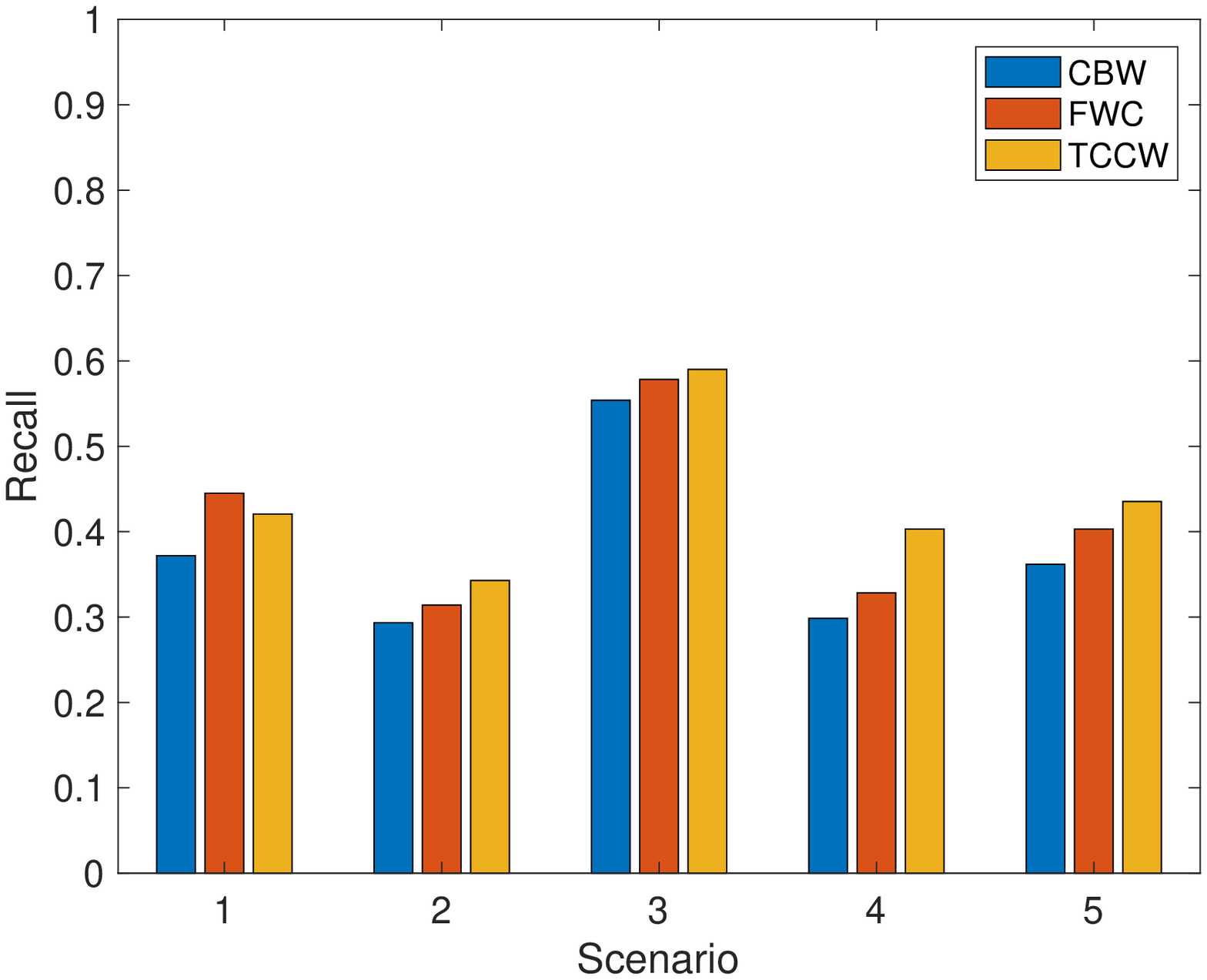}
    \end{minipage}}
\caption{System performance with different scenarios, difference of scenarios are listed in Table. \ref{table} }
\label{fig_experiment_scenario}
\end{figure*}

Comparing the superiority of distributed fog computing architecture to centralized cloud computing, we implement the system in both cloud computing and fog computing architecture.
{\color{blue} 
To compare the performance of our algorithm, we implement the collision system with three different algorithms as below:
}

\noindent
Cloud-Based Warning (CBW): The collision warning system is implemented in the centralized cloud computing architecture, specifically, vehicles upload their status information to a cloud server, which is far away from vehicles. In the experiment, we utilize a transmission latency obtained by a field testing experiment to simulate the communication delay between vehicles and cloud node. Cloud server predicts potential collision warning on the basis of vehicle status without trajectories calibration.

\noindent
Fog-Based Without Calibration (FWC): The collision warning system is implemented in the traditional fog computing architecture. In the fog computing architecture, the majority system is implemented as same as CBW, expects the network architecture. Due to new network architecture, communication latency drastically reduces. Vehicles upload their status to nearby fog nodes, and we use transmission delay obtained in real-world field testing as transmission latency of each packets. The fog node predicts the potential collisions without trajectory calibration.

\noindent
Trajectory Calibration based Collision Warning(TCCW): The details of TCCW are introduced in Section. \ref{Proposed Algorithm}. Fog node predicts potential collision by calibrating vehicle trajectory considering both transmission latency and packet loss.

\subsection{Experiment Result and Analysis}

Fig. \ref{fig_experiment_scenario} shows the precision and recall of three algorithms under different experiment scenarios.
It is obvious that the trajectories calibration improve the performance in both precision and recall in no matter what traffic conditions.
With the benefit of trajectory calibration in the fog node, vehicle status are much closer to the real-time status, in other words, vehicle positions with trajectories calibration are more closer to the real vehicle positions.

\begin{figure*}
\centering
    \subfigure[Precision with different headway]{
    \begin{minipage}{3.0in}
    \centering
    \label{fig_experiment_headway_one}
    \includegraphics[scale=0.40]{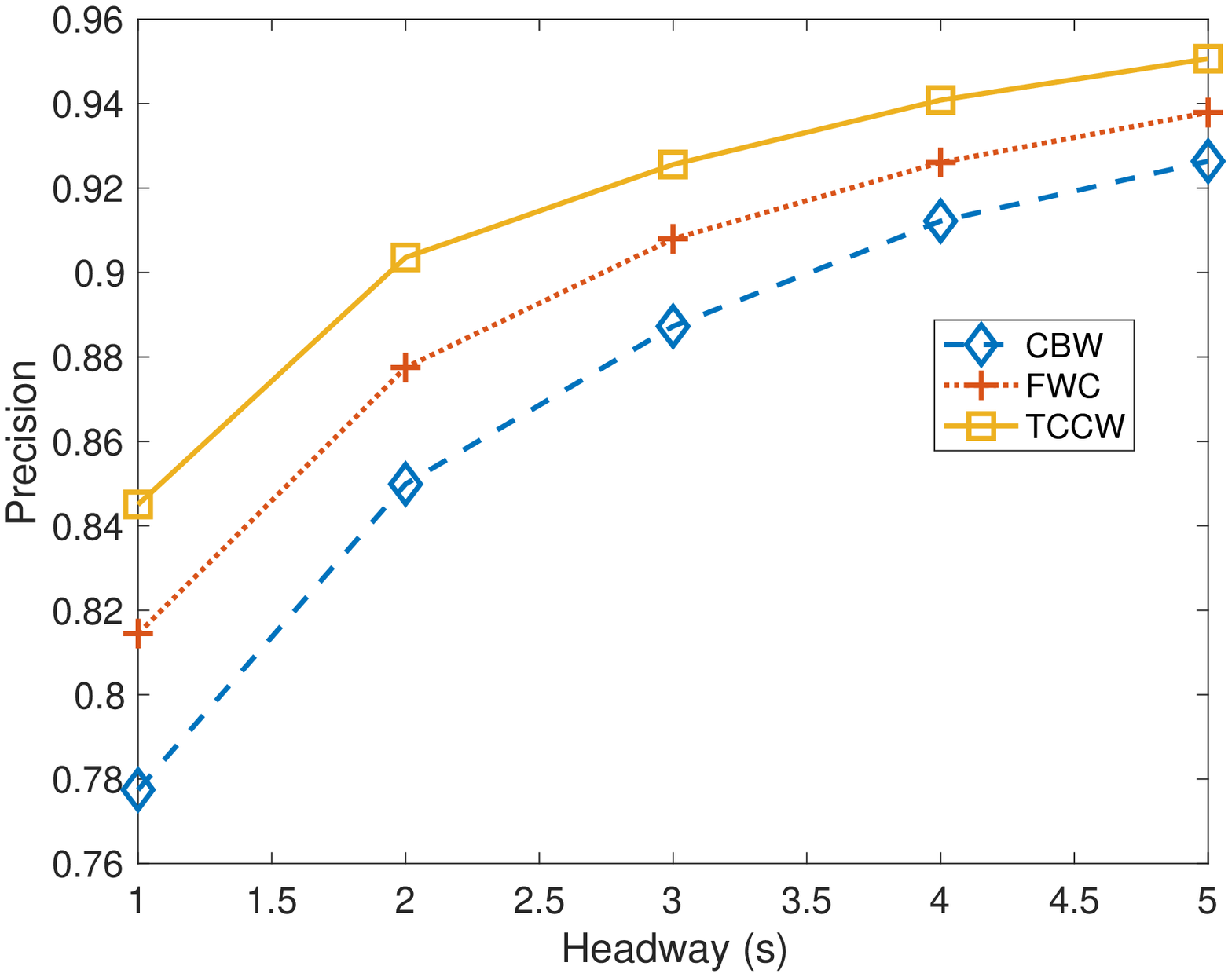}
    \end{minipage}}
    \subfigure[Recall with different headway]{
    \begin{minipage}{3.0in}
    \centering
    \label{fig_experiment_headway_two}
    \includegraphics[scale=0.40]{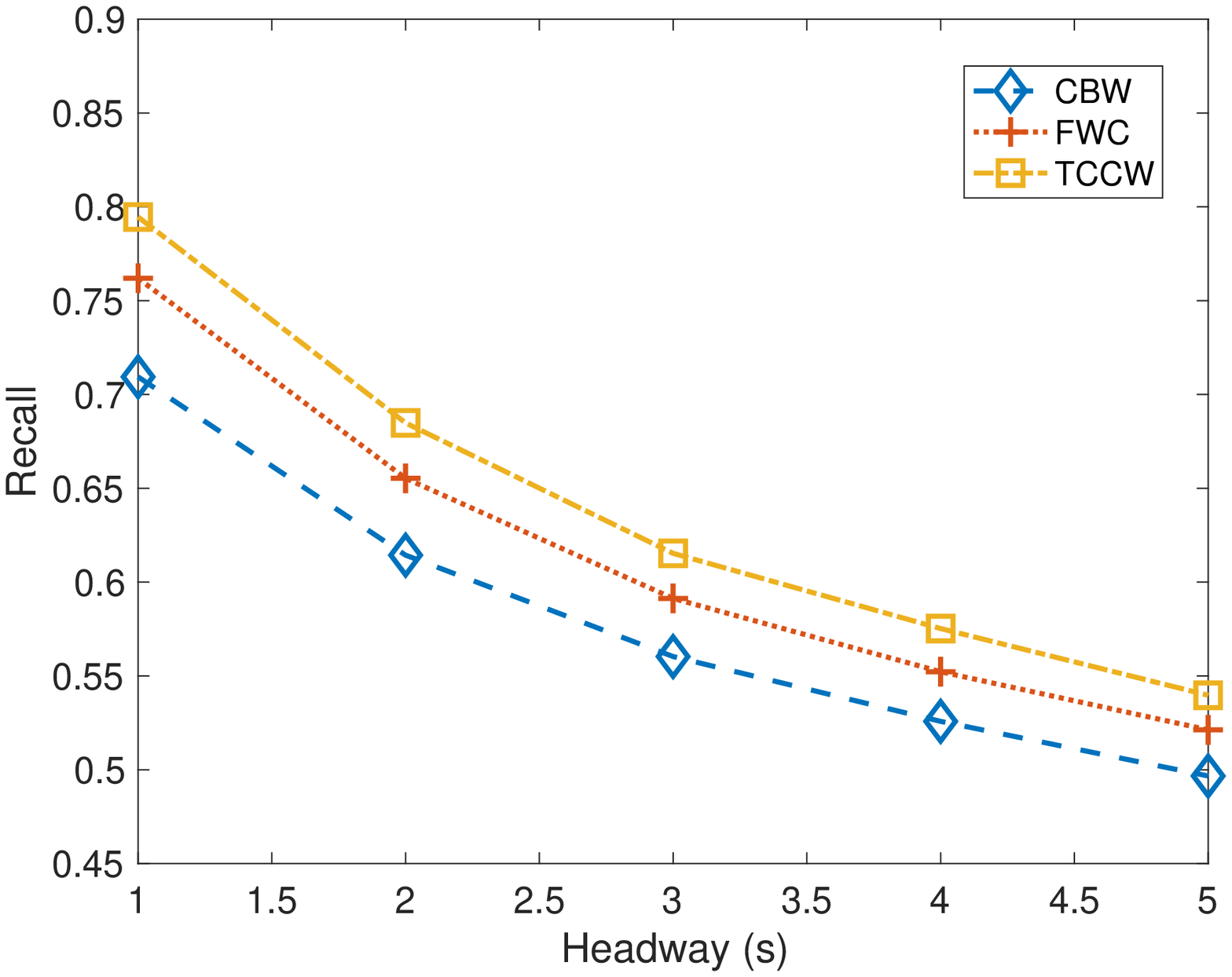}
    \end{minipage}}
\caption{System performance with different headways, which increases from 1 to 5 seconds}
\label{fig_experiment_headway}
\end{figure*}

\begin{figure*}
\centering
    \subfigure[Precision with different packet loss rate]{
    \begin{minipage}{3.0in}
    \centering
    \label{fig_experiment_plr_one}
    \includegraphics[scale=0.40]{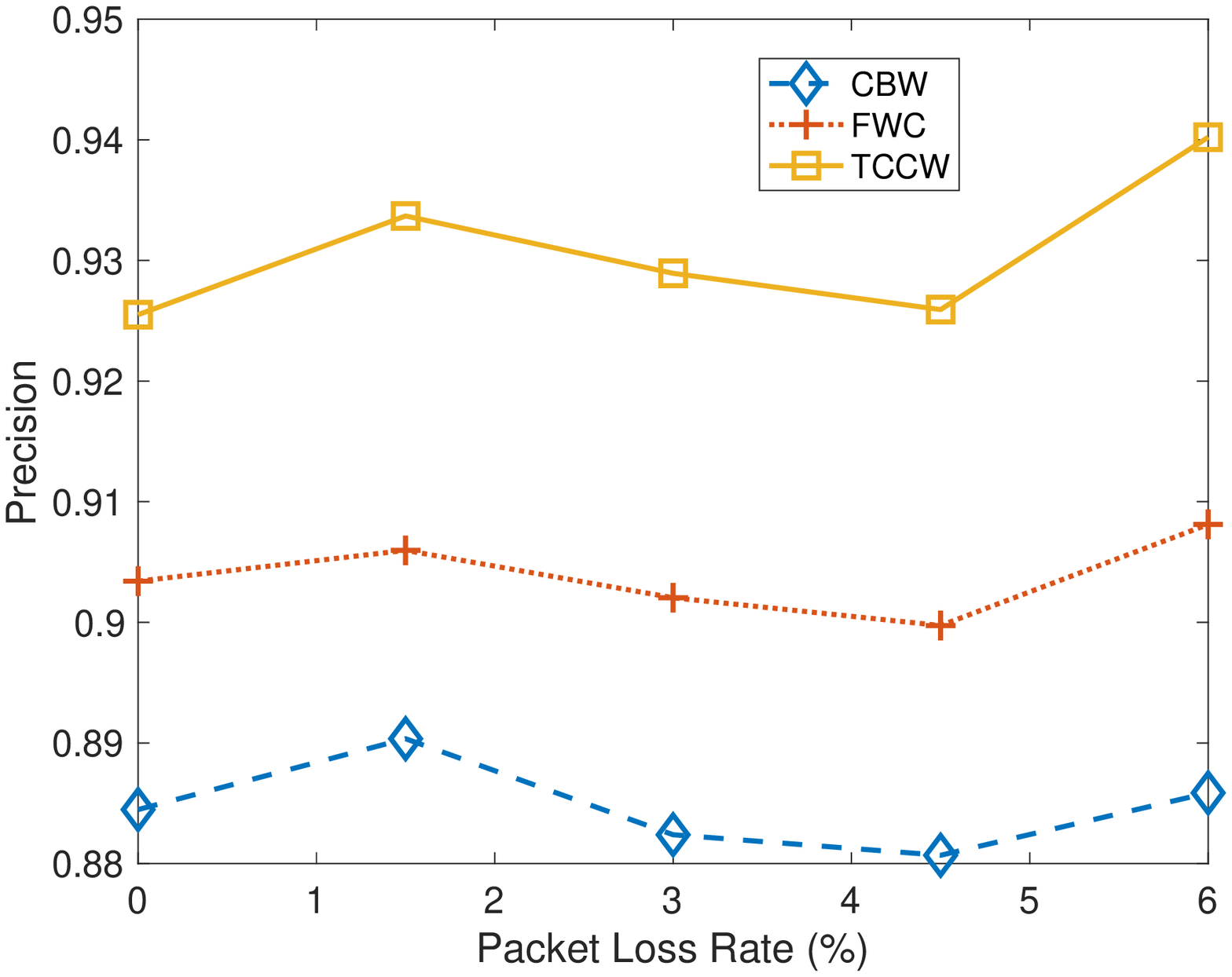}
    \end{minipage}}
    \subfigure[Recall with different packet loss rate]{
    \begin{minipage}{3.0in}
    \centering
    \label{fig_experiment_plr_two}
    \includegraphics[scale=0.40]{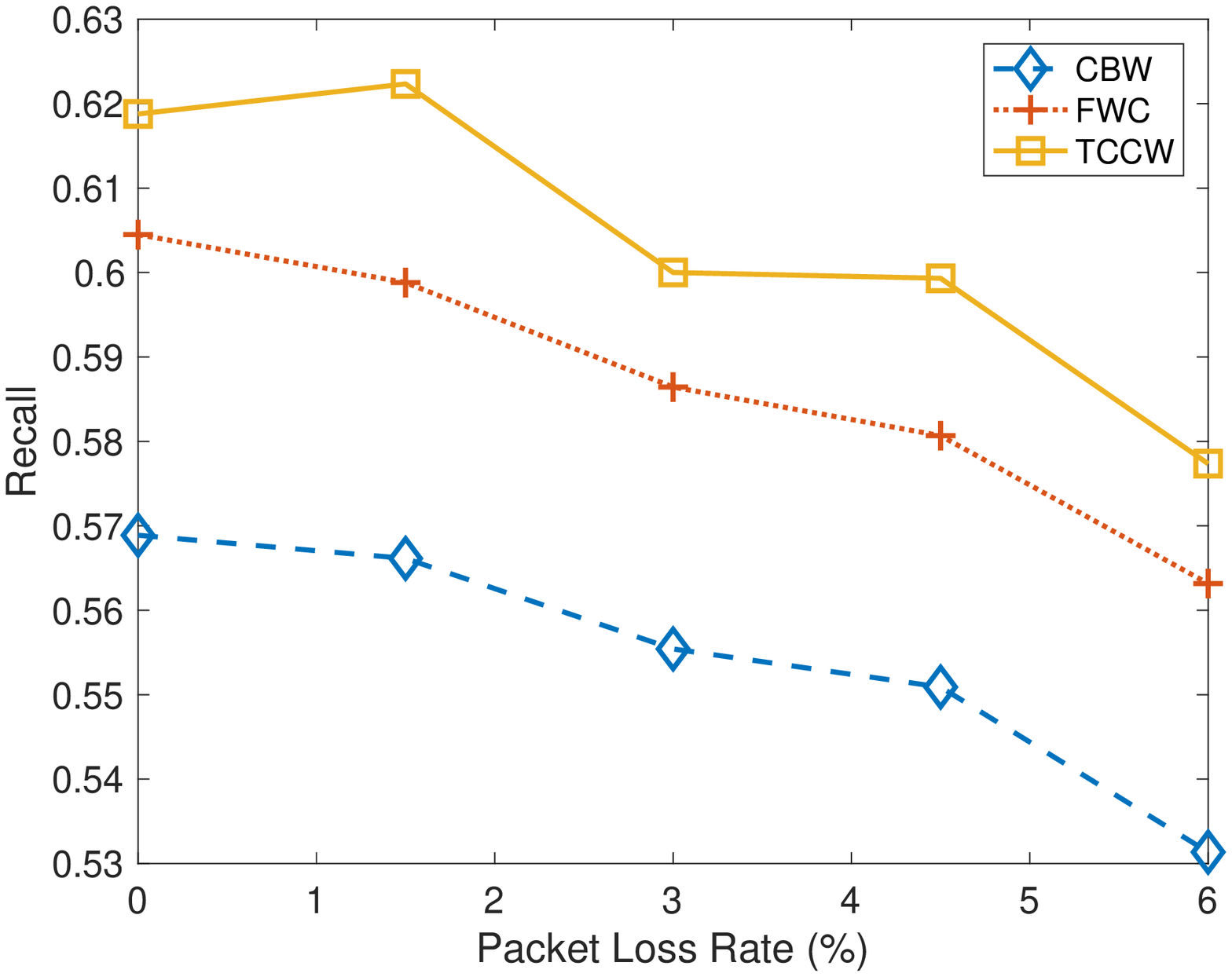}
    \end{minipage}}
\caption{System performance with different packet loss rate, which increases from 0\% to 6\%}
\label{fig_experiment_plr}
\end{figure*}

Fig. \ref{fig_experiment_headway} compares the precision and recall of three algorithms with different headway thresholds.
Fig. \ref{fig_experiment_headway} shows the headway threshold increases from 1 to 5 seconds. 
As the headway threshold increases, precision has improved.
Since headway threshold turns large, the number of predicted collisions get larger, in another word, $\left| W \right|$ get bigger.
Thus, the precision of proposed system is improved, on the other hand, recall reduces as headway threshold increased.

Fig. \ref{fig_experiment_plr} compares the precision and recall of three algorithms with different packet loss rates. 
As shown in Fig. \ref{fig_experiment_plr}, when the packet loss rate increases from 0\% up to 6\%, performance of CBW and FWC become worse, however, performance of TCCW is better than the two algorithms with the help of trajectory calibration.
The experiment result shows that due to the low communication latency supported by fog computing architecture, the performance of FWC is better than CBW. 
And we can also find that with the help of trajectory calibration, the performance of TCCW is better than both CBW and FWC.

\section{Conclusion}\label{Conclusion}
In this paper, we present a VFC based real-time collision warning architecture to support safety-critical ITS applications. 
We propose a TCCW algorithm, which calibrates vehicle trajectories by combining transmission latency estimation and packet loss detection.
We estimate the transmission latency by deriving a packet transmission delay fitting model in application-layer based on the Stable distribution, where the fitting data are obtained by implementing a DSRC based system in realistic field testing. 
Moreover, we detect the packet loss based on the historical information including data transmission frequency and locations of vehicles.
We give a comprehensive simulation and the result demonstrates the superiority of TCCW comparing with cloud-based warning and fog-based warning without trajectory calibration in terms of both recall and precision on collision warning.

\begin{acknowledgements}
    This work was supported in part by the National Natural Science Foundation of China under Grant No.61872049, No.61876025 and No. 61803054;  the Venture $\&$ Innovation Support Program for Chongqing Overseas Returnees (Project No. cx2018016), and the Fundamental Research Funds for the Central Universities (2019CDQYZDH030).
\end{acknowledgements}



\end{document}